\documentclass[12pt]{article}
\usepackage[centertags]{amsmath}
\usepackage{amsfonts} 
\usepackage{amssymb} 
\usepackage{amsthm}
\usepackage{graphicx}

\newcommand{\Z}{{\mathbb Z}}
\newcommand{\R}{{\mathbb R}}

\newcommand{\be}{\begin{equation}}
\newcommand{\eeq}{\end{equation}}
\newcommand{\bea}{\begin{eqnarray}}
\newcommand{\eea}{\end{eqnarray}}
\newcommand{\ba}{\begin{array}}
\newcommand{\ea}{\end{array}}
\def\nn{\nonumber}
\newcommand{\ft}[2]{{\textstyle\frac{#1}{#2}}}

\newcommand{\ii}{\mathrm{i}}
\newcommand{\ee}{\end{equation} }

\newcommand{\cM}{{\mathcal{M}}}

\newcommand{\one}{{\rm 1\kern -.9mm l}}

\newdimen\tableauside\tableauside=1.0ex
\newdimen\tableaurule\tableaurule=0.4pt
\newdimen\tableaustep

\def\phantomhrule#1{\hbox{\vbox to0pt{\hrule height\tableaurule
width#1\vss}}}
\def\phantomvrule#1{\vbox{\hbox to0pt{\vrule width\tableaurule
height#1\hss}}}
\def\sqr{\vbox{%
 \phantomhrule\tableaustep
\hbox{\phantomvrule\tableaustep\kern\tableaustep\phantomvrule\tableaustep}%
 \hbox{\vbox{\phantomhrule\tableauside}\kern-\tableaurule}}}
\def\squares#1{\hbox{\count0=#1\noindent\loop\sqr
 \advance\count0 by-1 \ifnum\count0>0\repeat}}
\def\tableau#1{\vcenter{\offinterlineskip
 \tableaustep=\tableauside\advance\tableaustep by-\tableaurule
 \kern\normallineskip\hbox
   {\kern\normallineskip\vbox
     {\gettableau#1 0 }%
    \kern\normallineskip\kern\tableaurule}%
 \kern\normallineskip\kern\tableaurule}}
\def\gettableau#1 {\ifnum#1=0\let\next=\null\else
 \squares{#1}\let\next=\gettableau\fi\next}
\newcommand{\Yfund}{\tableau{1}}
\newcommand{\Ysymm}{\tableau{2}}
\newcommand{\Yasymm}{\tableau{1 1}}

\newcommand{\Smod}{S_{\rm mod}}

\begin{document}
\begin{titlepage}
\begin{flushright}
{ROM2F/2011/06}\\
\end{flushright}
\begin{center}
{\large \sc Multi instanton tests of holography }\\
\vspace{1.0cm}
{\bf F.Fucito}, {\bf J. F. Morales},  {\bf D. Ricci Pacifici}\\
 I.N.F.N. Sezione di Roma Tor Vergata\\ and \\{\sl Dipartimento di Fisica, Universit\'a di Roma ``Tor Vergata''\\
Via della Ricerca Scientifica, 00133 Roma, Italy}\\
\end{center}
\vskip 2.0cm
\begin{center}
{\Large \sc }
\end{center}

\abstract{Gauge theories living on stacks of D7-branes  are holographically related to 
IIB gravitational backgrounds with a varying axion-dilaton field (F-theory). The axion-dilaton
field is generated by D7, O7 and D-instanton sources and can be written in terms of the chiral correlators of the eight dimensional 
gauge theory living on the D7-branes.  
  Using localization techniques, we prove that the same correlators determine the gauge
  coupling of the four-dimensional ${\cal N}=2$ supersymmetric $SU(2)$ gauge theories living on the elementary D3-brane which probes the 
F-theory geometries.  
   }
\end{titlepage}

\tableofcontents

\section{Introduction and Summary}
\label{sec:intro}

F-theory provides us with some of the few known explicit examples of  non-conformal gauge gravity 
duals at the non-perturbative level. The simplest setting involves a stack of  $N$ D7-branes on top of an O7-plane in 
flat ten-dimensional spacetime\footnote{This local brane system arises in type I' theory  
   (the theory obtained by T-dualizing a $T^2$ torus of  type I theory on $T^2$) after locating 
   $N$ D7-branes in the neighborhood of one of the four O7 planes.}.  The D7 worldvolume theory is described by a maximally supersymmetric gauge  theory    in eight space time dimensions  with gauge group $SO(2N)$.  The gravity side is given by F-theory at a $D_N$-singularity or equivalently
  by a type IIB orientifold with varying axion-dilaton field, $\tau(z)$, with $z$ the coordinate on the complex plane  transverse to the D7 branes.
   
 The $SO(8)$ case, with $N=4$ D7-branes on top of the O7-plane, is special because 
  the tadpoles are exactly canceled  and the axion-dilaton field, $\tau_0$, is constant along the $z$-plane. Taking the D7-branes away from the O7 plane, let us say at positions $a_u$,  generates a non trivial dependence $\tau(z,\tau_0,a_u)$ for the axion-dilaton  
field. Relying on the $SO(8)$
  symmetries of the background, it was proposed in \cite{Sen:1996vd} to identify  this function  with  the gauge coupling $\tau_{\rm gauge}(z,\tau_0,m_u)$ described by the Seiberg-Witten curve of a ${\cal N}=2$ supersymmetric 
  $SU(2)$ gauge theory with four hypermultiplets of masses $m_u =a_u$, transforming in the fundamental representation of the gauge group \cite{Seiberg:1994rs}. 
  $z$ parametrizes the Coulomb branch of the moduli space of the gauge theory and $\tau_0$ is the UV coupling.
  This proposal was further supported by the observation in \cite{Banks:1996nj} that the same gauge theory describes the dynamics of a D3-brane 
probe of the F-theory geometry. 
 
More recently in \cite{Billo':2011uc}, the $SO(8)$ F-theory background was  explicitly 
derived from string diagrams describing the rate of emission of the axion-dilaton field from
D7, O7 and D(-1) sources. The results were written in the suggestive form \cite{Billo:2010mg}
 \be
\tau_{\rm sugra} (z,\tau_0,a) =\langle  {\cal O}_\tau(\Phi)  \rangle_{D7}\qquad ~~~~~~~~a =\langle \Phi \rangle_{D7}\label{strik}
\ee
where
\be
{\cal O}_\tau(a)  =\tau_0+  \ln {\rm det} \left(1-{a \over z} \right) 
 \label{otau}
\ee
is the operator in the eight space time dimensional gauge theory
dual to the axion-dilaton field on the gravity side.  $a$ is an $8\times 8$ 
antisymmetric matrix with purely imaginary eigenvalues $\pm a_u$ parametrizing the D7-brane
positions.
The  operator ${\cal O}_\tau$ is determined by 
disk amplitudes involving an insertion of a $\tau$-field and any number of open string vertices. 
 We remark that the correlator on the r.h.s. of (\ref{strik}) receives an 
infinite tower of D(-1)-instanton corrections resulting into an infinite series in $e^{2\pi i \tau_0}$ for the axion-dilaton field.
   
Holography requires that the same function  describes
 the gauge coupling $\tau_{\rm gauge} $ on the
  four space time dimensional theory living on the D3-brane probe
 \be
\tau_{\rm gauge} (z,\tau_0,a) =\langle  {\cal O}_\tau(\Phi)  \rangle_{D7}   \label{strik2}
\ee 
   This is consistent with the fact that, according to (\ref{otau}),  ${\cal O}_\tau(a)$ is nothing but
 the perturbative contribution (tree and one-loop ) to the gauge coupling on the 
D3-brane probe.  D-instantons modify both sides of (\ref{strik2}) in a rather asymmetric way
  and therefore an agreement at the non-perturbative level is far from obvious. In particular,
  exotic instantons in the eight dimensional theory (the right hand side) are entangled with the ``standard'' gauge instantons in four space time dimensions (the left hand side).

 The intriguing relation (\ref{strik2})  was verified  in \cite{Billo:2010mg} for the first few instanton orders by an explicit comparison of the  results coming from the corresponding Seiberg-Witten curve against those for  the eight space time dimensional chiral  correlators computed in \cite{Fucito:2009rs}.   The aim of this paper is to prove that this relation 
    holds at any instanton order. The proof will rely on 
   multi instanton formulae  obtained via localization  \cite{Nekrasov:2002qd,Bruzzo:2002xf}. 
   The analysis will be extended to a large class of F-theory backgrounds
 with sixteen and eight supersymmetric charges. More precisely, we will  consider systems  made of $N$ D7 branes and an O7 plane on $\R^{10}$ or $\R^6\times \R^4/\Z_2$. The associated gauge symmetries of the D7-worldvolume theories are
$SO(2N)$ and $U(N)$ respectively.  Both eight-dimensional backgrounds will be tested
via elementary D3-branes supporting a ${\cal N}=2$ supersymmetric $SU(2)$ gauge theories with
$N$ fundamentals. The instanton moduli spaces for these brane systems were worked out in \cite{Fucito:2009rs,Billo:2009di,Billo':2010bd}. 
       
    Finally we remark that the relation (\ref{strik2}) holds even in the extreme case $N=0$, where no D7-branes are included in the F-theory background.  The $\tau_{\rm sugra}$ in this case corresponds to the axion-dilaton field generated by  a single O7-plane.  On the other hand $\tau_{\rm gauge}$ describes the gauge coupling of the pure ${\cal N}=2$ supersymmetric $SU(2)$ gauge theory living on a D3-brane probe  of this geometry. We stress that even in this simple case one finds a non-trivial  F-theory elliptic fibration incorporating the effects of D-instantons on the geometry. 
 This case can be thought of as the decoupling limit where the masses of all the fundamentals are sent to
 infinity by bringing the D7-branes far away from the O7 plane. 
 More interestingly, the correlator on the right hand side of (\ref{strik}) describes well defined F-theory 
 geometries also in the case with $N>4$ D7-branes where the $SU(2)$ gauge theory in the probe is not asymptotically free and a Seiberg-Witten type analysis is not available. It would be nice to understand what is the geometry underlying such F-theory backgrounds.

\section{$SO(2N)$ models}

We start by considering a system of D(-1), D3 and D7 branes in presence of an O7 plane 
in a flat ten dimensional space. 
The D(-1)-branes or D-instantons can be thought of either as gauge instantons on the D3-brane or as exotic instantons on the eight dimensional gauge theory living 
on the D7-branes. In each  case, the instanton moduli are associated to the massless modes of the open strings connecting the various branes. 

The background is specified by  the positions, $a_u$, of the D7-branes in the tranverse plane 
which parametrize the vevs of the chiral  
scalar field $\Phi$  in the eight dimensional
worldvolume theory. More precisely, using the $SO(2N)$ invariance one can always bring the antisymmetric matrix
$\langle \Phi \rangle_{\rm cl}$ into the diagonal form   
\be
\langle \Phi \rangle_{\rm cl}={\rm diag} \{ a_u, -  a_u  \}
\ee
with $a_u$ purely imaginary, $u=1,\ldots N$.
 At the quantum level the background is modified by the presence of the D-instantons and it is better described by the eight dimensional prepotential 
${\cal F}(\Phi)$ and the chiral
correlators $\langle {\rm tr} \Phi^J \rangle_{D7}$. The prepotential and chiral correlators 
  are computed by integrals over the multi-instanton moduli  space \cite{Fucito:2009rs,Billo:2009di}. 
  After suitable deformations that regularize the eight dimensional spacetime volume 
  ${\rm vol_{D7}}\sim {1\over \epsilon_1\epsilon_2\epsilon_3\epsilon_4}$,  these integrals localize around a finite number 
of points where they can be explicitly evaluated. More precisely,
  the integrals over the multi-instanton moduli space reduce to contour integrals on $\chi_i$
  (the positions of the D(-1)-branes) of  a super-determinant $ {\rm det} Q^2$,  with   
  $Q$ an equivariantly deformed BRST charge. The parameters $\epsilon_\ell, a_u, \chi_i$ together with
  the positions $b_m$ of the D3-brane probes parametrize the Lorentz and gauge symmetries
  broken by the instantons and determine the eigenvalues of
   $Q^2$. 
  
  \subsection{The D(-1)D7 system}
  
  We start by considering the D(-1)D7 system  with no D3-branes.  
The theory on the D7-branes lives in eight dimensions and carries a gauge group $SO(2N)$.
Instantons are realized by the inclusion of $K$ fractional D(-1)-branes with symmetry group $SO(K)$.  
In the following we will use the notations employed in \cite{Fucito:2009rs} to which we refer for further details.
 
    The $K$-instanton partition function of the D7 worldvolume theory can be written as 
 \be
 Z_{D7,K}=\int d\cM_{K}  ~e^{-\Smod(\cM_{K})}=\int \prod_{i=1}^k \frac{d{\chi_i}}{2\pi\ii} \,
 w_{D7} (\chi,a)  \label{zd7}
 \ee
  with $w_{D7} (\chi)$ the contributions of D(-1)D(-1) and D(-1)D7 open strings
  to the determinant of $Q^2$. 
  The matrix $\chi$ is $K\times K$ antisymmetric with purely imaginary eigenvalues 
  $\chi_I$ specifying the positions of the instantons and their images along the plane 
  transverse to D7-branes. We write 
  \be
\chi_I =\left\{     
\begin{array}{lll}
 (\chi_i,-\chi_i) &  {\rm for} & SO(2k)  \\
  (\chi_i,-\chi_i,0) &  {\rm for} & SO(2k+1)  \\\end{array}
\right.   \qquad    i=1,\ldots k   \label{chis}
\ee  
   depending on whether $K$ is even or odd. We denote by $k=[\ft{K}{2}]$ the number of regular instantons with positions $\chi_i$. 
We notice that  for $K$ odd a fractional instanton is always stuck at the origin. 
      The integral over  $\chi_i$  can be computed as a contour integral
  on the upper half plane after taking the poles prescription $1>>{\rm Im} \epsilon_1 >> {\rm Im} \epsilon_2 >> {\rm Im} \epsilon_3>>{\rm Im} \epsilon_4 >>  {\rm Im} \,a_u$. In addition we require
  \be
  \epsilon_1+\epsilon_2+\epsilon_3+\epsilon_4=0  \label{sume}
  \ee
  in order to ensure the invariance of the BRST charge $Q$.
  The integrand in (\ref{zd7}) can be written as \cite{Fucito:2009rs}
  \bea 
w_{D7} \,(\chi,a) &=&{ c_K\over k! }
\prod_{\ell=1}^4 \prod_{I,J}^K {}'  \left[ {  ( \chi_{IJ}+s_\ell) \over 
( \chi_{IJ}+\epsilon_\ell) } \right]^{1\over 2}
\prod_{\ell=1}^4 \prod_{I=1}^K    \left[{  1 \over 
( 2\chi_{I}+\epsilon_\ell) ( 2\chi_{I}+s_\ell)} \right]^{1\over 2}\nn\\
&& \times \prod_{I=1}^K   \prod_{u=1}^{\bf N} (\chi_I- a_u)   \label{wk1}
\eea
  with $\chi_{IJ}=\chi_I-\chi_J$ and
  \be
  s_1=0 \qquad s_2=\epsilon_1+\epsilon_2  \qquad s_3=\epsilon_1+\epsilon_3
   \qquad s_4=\epsilon_2+\epsilon_3
  \ee 
  The prime in (\ref{wk1}) denotes the omission of the zero eigenvalues $(\chi_{II}+s_1)$ from the product. 
 The three products in (\ref{wk1}) correspond to the contributions of D(-1)D(-1), D(-1)O7
 and D(-1)D7 open strings respectively. In particular the denominator of the first product comes from 
 the four complex matrices $B_\ell$, $\ell=1,..4$ specifying the positions of the instantons along the 
   D7 worldvolume while the numerator accounts for the generalized ADHM contraints $D_\ell$ 
defining the instanton moduli space. The second product implements
   the projections of $B_\ell$ and $D_\ell$ into the symmetric and adjoint representations
   of $SO(K)$ respectively. Finally the last product accounts for the D(-1)D7 strings 
  with half the degrees of freedom of a bifundamental of $SO(K)\times SO(2N)$. 
   Plugging (\ref{chis}) into (\ref{wk1}) one can see that all
     square roots cancel out as expected. 
       Finally $c_K=(-)^{k+1}(2)^{k-K }$ are numerical coefficients\footnote{They can be determined requiring that ${\cal F}_{D7}$ defined below
      is finite in the limit $\epsilon_\ell\to 0$.} . 
     
  The prepotential is defined as
  \be
  {\cal F}_{D7}=-\lim_{\epsilon_\ell \to 0}\epsilon_1\epsilon_2\epsilon_3\epsilon_4 \, \ln Z_{D7}(q) 
  \label{ff4}
  \ee
   with 
   \be
   Z_{D7}(q)=\sum_{K=0}^\infty Z_{D7,K}\, q^K
   \ee
   and $Z_{D7,0}=1$. The chiral correlators are given in terms
   of the same integrals with extra $\chi$-insertions according to \cite{Fucito:2009rs}
   \be
  \langle {\rm tr} e^{z \Phi} \rangle_{ {\rm inst} } = 
 {1\over Z_{D7} } \prod_{\ell=1}^4 (1-e^{z \epsilon_\ell})
\sum_{K=1}^\infty q^K  \int \prod_{i=1}^k \frac{d{\chi_i}}{2\pi\ii} \,
  w_{D7} (\chi,a)\, {\rm tr} e^{ z \chi}    \label{trphi}
  \ee
     In the limit $\epsilon_\ell\to 0$ (\ref{trphi}) reduces to
  \be
  \langle {\rm tr} e^{ z \Phi} \rangle_{ {\rm inst}}    =\lim_{\epsilon_\ell\to 0}  
 { \epsilon_1\epsilon_2\epsilon_3\epsilon_4 \over Z_{D7} } 
 \sum_{K=1}^\infty q^K \int \prod_{i=1}^k \frac{d{\chi_i}}{2\pi\ii} \,
  w_{D7} (\chi,a)\, z^4\, {\rm tr} e^{ z \chi}
  \label{chiral}
 \ee
 or equivalently
 \be
  \left\langle {\rm tr} {\Phi^{J+4}\over (J+4)!} \right\rangle_{ {\rm inst}}  =  \lim_{\epsilon_\ell\to 0}  
 { \epsilon_1\epsilon_2\epsilon_3\epsilon_4 \over Z_{D7} } 
  \sum_{K=1}^\infty q^K  \int \prod_{i=1}^k \frac{d{\chi_i}}{2\pi\ii} \,
  w_{D7} (\chi,a)\, {\rm tr} { \chi^{J}\over J!}
  \label{chiral}
 \ee
   with  $\langle {\rm tr} \Phi^{J} \rangle_{ {\rm inst}}=0$ for $J<4$. 
   We remark that the correlator
   in the right hand side of (\ref{chiral}) is defined even in the case $N=0$ (no D7-branes) where
   the left hand side of this equation looses its sense. In this case, we will loosely keep the notation 
    $ \left\langle {\rm tr} \Phi^{J}  \right\rangle_{ {\rm inst}} $ as a shorthand  for the 
    associated correlator on the right hand side of (\ref{chiral}).

    \subsection{The coupling on the D3-brane probe}
  
 Now we introduce a stack of M D3-branes probing the D7-background geometry.
  The theory living on the D3-branes is ${\cal N}=2$ supersymmetric with gauge group $Sp(M)$
and flavor group $SO(2N)$\footnote{In our conventions $Sp(1)=SU(2)$.}. The 
$Sp(M)$  matter content  is given by one  hypermultiplet transforming in the antisymmetric representation and $N$ hypermultiplets transforming in the fundamental representation of the gauge field.

The beta function coefficient reads
\footnote{We use the conventions: 
$
b= 2 \Big[T({\rm adj}) - \sum_r n_r T(r)\Big] $ 
with $T(r)$ the index of the representation $r$ of $Sp(M)$ given by
 $T({\rm adj})=  \frac{2M+ 2}{2} $, 
$T(\Yasymm)= \frac{2M-2}{2}$  and
$T(\Yfund)=\frac{1}{2}$.} 
  \bea
 b_{Sp(M)} &=&  4-N
 \eea
 The case $N=4$ is special in the sense that the four dimensional gauge theory is conformal for any $M$. 

Now let us consider the effects of D-instantons. The $K$-instanton partition function is given by
 \be
 Z_{D3,K} =\int \prod_{i=1}^k \frac{d{\chi_i}}{2\pi\ii} \,
 w_{D7} (\chi,a)\,w_{D3} (\chi,b)   
 \label{zd3}
 \ee
  with   $w_{D7}$ given by (\ref{wk1}) and  
  \bea 
w_{D3} \,(\chi,b) &=& 
\prod_{m=1}^M \prod_{I=1}^K  {  ( \chi_{I}-b_m)^2 -\left(   
\ft{\epsilon_3-\epsilon_4}{2} \right)^2 
 \over   ( \chi_{I}-b_m)^2 -\left(\ft{\epsilon_1+\epsilon_2}{2}\right)^2}    \label{wd3}
\eea
giving the contributions of D(-1)D3 strings. 
  The four dimensional prepotential is defined by \cite{Billo':2010bd}
  \be
  {\cal F}_{D3}(a_u,b_m)=  -\lim_{\epsilon_{\ell} \to 0}\epsilon_1\epsilon_2 
      \ln   \left( {Z_{D3}(a_u,b_m) \over Z_{D7}(a_u) }   \right) 
   \label{fd3}
  \ee
   The term  $\epsilon_1\epsilon_2 \ln Z_{D7}=- {1\over \epsilon_3\epsilon_4} {\cal F}_{D7}$ subtracts a $b_m$-independent divergent 
term of eight-dimensional origin and
leaves a finite result for the four dimensional prepotential ${\cal F}_{D3}$.
    
        The gauge coupling on the D3-probe is defined by
    \be
2\pi i  \, \tau^{\rm inst}_{mn}=   {\partial^2 {\cal F}_{D3}\over \partial b_m \partial b_n}  \label{taumn}
    \ee  
  In the rest of this section we will show that  
$\tau_{mn}$  can be written entirely in terms of the chiral correlators of the D7 brane gauge theory . 
 
 First plugging (\ref{fd3}) into (\ref{taumn}) and using the fact that $Z_{D7}$ does not 
 depend on $b_m$ one finds
   \be
2\pi i  \,  \tau^{\rm inst}_{ mn} =  \lim_{\epsilon_{\ell} \to 0}\epsilon_1 \epsilon_2 \left[ -{1\over  Z_{D3}} {\partial^2 Z_{D3} \over \partial b_m \partial b_n}+{1\over  Z_{D3}^2} {\partial Z_{D3} \over \partial b_m }{\partial Z_{D3} \over \partial b_n }  \right]
  \label{tau1}
\ee
 Now let us evaluate the right hand side of (\ref{tau1}) in the limit $\epsilon_\ell \to 0$ limit. 
 Taylor expanding $w_{D3}$ in (\ref{wd3}) in the previous limit and for large $b_m$, leads to 
\be
w_{D3} \,(\chi,b)=1+ \epsilon_3 \epsilon_4 \sum_{J=0}^\infty \sum_{m=1}^M   
{{\rm tr} \chi^{J} \over b_m^{J+2}   }  (J+1)+\ldots    \label{wd32}
\ee
where (\ref{sume}) was used. The behavior $w_{D3} \,(\chi,b) \approx 1$ follows from the fact that
the D3-D3 sector is effectively  ${\cal N}=4$ supersymmetric  and therefore has trivial instanton 
corrections to the four dimensional prepotential. 
The ${\cal N}=4$ supersymmetry  is broken by the D7-branes which introduce fundamental 
matter and by the O7-plane which projects the ${\cal N}=2$ vector multiplet and the hypermultiplet
into different representations of the gauge group. These contributions are encoded in the $w_{D7} \,(\chi,a)$  term in (\ref{wk1}).
 
Plugging (\ref{wd32}) into (\ref{zd3}),  one sees that only the first term in (\ref{tau1}) contributes at order  $\epsilon_3\epsilon_4$ and one finds $\tau^{\rm inst}_{mn}=\delta_{mn} \tau^{\rm inst}_m$
with
\bea
&& 2\pi i  \, \tau^{\rm inst}_m =- \lim_{\epsilon_{\ell} \to 0} { \epsilon_1 \epsilon_2 \epsilon_3 \epsilon_4 \over Z_{D7} }
  \sum_{K=1}^\infty \sum_{J=0}^\infty   \, q^K \, \ft{(J+3)!}{J! }\int \prod_{i=1}^k \frac{d{\chi_i}}{2\pi\ii} \,
 w_{D7} (\chi,a)\,   
{ {\rm tr}   \chi^{J} \over b_m^{J+4} }  \nn\\
&&= -  \sum_{J=1}^\infty    {1  \over  J b_m^{J} } \left \langle  
{      {\rm tr} \,  \Phi^{J}  } \right  \rangle_{ {\rm inst}} 
=   \left \langle   \ln {\rm det} \left(1-{\Phi\over b_m} \right) 
  \right  \rangle_{ {\rm inst}}    \label{taumn2}
\eea
 The result (\ref{taumn2}) shows that the gauge coupling matrix of the $Sp(M)$ gauge theory factorizes, in the limit $\epsilon_\ell \to 0$, into $M$ 
copies of a 
 $Sp(1)\sim SU(2)$ gauge theories, each copy coming with $N$ massive flavors transforming in the fundamental representation of $SU(2)$. Without loosing 
generality we can then set $M=1$ and rename $\tau_m\to \tau$, $b_m\to b$. 
Moreover, the result (\ref{taumn2}) shows that  the gauge coupling $\tau$    
 can be written as an infinite sum of chiral correlators in the eight dimensional flavor gauge theory!
 The   correlators  $\left \langle  
      {\rm tr} \,  \Phi^{J}   \right  \rangle_{ {\rm inst}} $ are given by the moduli space integrals (\ref{chiral}). Explicit results  up to $k=3$  
are  collected in the Appendix.

 \subsubsection{Explicit results up to $k=3$}

In this section we display some explicit formulas for the first few instanton contributions
to $\tau_{\rm inst}$. Plugging  (\ref{8corr})  into (\ref{taumn2}) one finds
     \bea
  \tau_{\rm inst}(b)  &&=
    - \sum_{J=0}^{12}  {1\over J} \left\langle {\rm tr} \frac{\Phi^{J}}{b^J} \right\rangle_{\rm inst} \label{taub}      \\
    &&=
     \frac{3 \, q \, \sqrt{{A}_{N}}}{{b}^{4}}+{q}^{2} \, \left ( \frac{3 \, {A}_{N-2}}{32 \, {b}^{4}}-\frac{15 \, {A}_{N-1}}{16 \, {b}^{6}}+\frac{\hbox{105} \, {A}_{N}}{32 \, {b}^{8}}\right ) \nn\\
&&  +   {q}^{3} \, \sqrt{{A}_{N}} \, \left ( \frac{3 \, {A}_{N-4}}{64 \, {b}^{4}}-\frac{5 \, {A}_{N-3}}{16 \, {b}^{6}}+\frac{35 \, {A}_{N-2}}{32 \, {b}^{8}}-\frac{21 \, {A}_{N-1}}{8 \, {b}^{10}}+\frac{\hbox{165} \, {A}_{N}}{32 \, {b}^{12}}\right ) \nn
     \eea
     with  $A_m$, $m\,=\,1,\ldots,N$ a basis for the Casimirs of $SO(2N)$
     given  by
     \bea
     A_s \,=\, \sum_{i_1<i_2\ldots<i_s} a_{i_1}^2\ldots a_{i_s}^2~~~~~~~~~~~~\nn\\ 
      A_N \,=\,  a_1^2 \ldots a_N^2 ~~~~~~~~~A_0 \,=\, 1~~~~~~~~~~A_{s<0} \,=\, 0  \label{ans}
    \eea
 Formula (\ref{taub}) determines the axion-dilaton field generated by a stack of $N$ D7-branes.   
For $N\leq 4$ it matches the gauge coupling coming from the Seiberg-Witten curve 
associated to a four dimensional theory with gauge group SU(2) and N fundamental hypermultiplets.
Explicitly  the  curves for $N<4$ can be written as
     \be
     y^2+R(x) y+q=0  \label{swcurve}
     \ee  
     with 
      \be
R(x)={x^2-e^2\over \sqrt{\prod_{u=1}^N (x-a_u) }}  
    \label{lambda}
 \ee
and 
\be
\lambda= {dz\over 2 \pi i }   {z R'(z) \over \sqrt{R(z)^2-4 q}  }
\ee
the Seiberg-Witten differential. $b$ and ${\partial {\cal F}\over \partial b}$ are defined by the 
two periods
of the differential and $\tau={\partial^2 {\cal F}\over \partial b^2}$
\footnote{Alternatively the prepotential from the Matone relation $2 e^2=2 q {dF\over dq} +2 a^2$. }. 
 The conformal case  $N=4$ was
  worked out  in \cite{Billo:2010mg}. 
 
We remark that the F-theory axion-dilaton result (\ref{taub}) is well defined even in cases 
like  $N=0$ where  the eight-dimensional gauge theory is missing or $N>4$ where the four-dimensonal gauge theory on the D3-brane probe is not asymptotically free and curves of the Seiberg-Witten type  are not available.
The former case corresponds to an F-theory background made out of an O7-plane
   and no D7-branes.  
   The eight-dimensional amplitude in this case follows from (\ref{taub}) after setting $N=0$. 
   We recall that in the case of no D7-branes  the notation $\langle {\rm tr} \Phi^{J} \rangle_{\rm inst}$ is a shorthand for the D-instanton correlator on the right hand side of (\ref{chiral}). 
 Up to the third order in q one finds
     \be
 \tau_{\rm inst}(b)= - \sum_{J=0}^{12}  {1\over J}\left \langle {\rm tr} \frac{\Phi^{J}}{b^J}\right \rangle_{\rm inst}  =\, 3\frac{ \, q}{ \, {b}^{4}}+ \frac{\hbox{105}}{32}\frac{ \, {q}^{2}}{ \, {b}^{8}}
   +   \frac{\hbox{165}}{32}\frac{ \, {q}^{3}}{ \, {b}^{12}}  
     \ee
      which reproduces the first instanton corrections to the  gauge coupling of the pure $SU(2)$ gauge theory as claimed.  The F-theory background generated by O7, D(-1) branes 
   is then described by  the pure $SU(2)$ Seiberg- Witten curve given by 
   (\ref{swcurve}) with $N=0$.

\section{ $U(N)$ models}

The results we found in last section can be easily generalized to F-theory backgrounds
 with less supersymmetries  and different gauge groups. 
 Consider the same brane system on   $\R^6\times \R^4/\Z_2$. In general $\Z_2$
 acts on the Chan-Paton indices and projects the brane symmetry groups into
 unitary or  SO/Sp gauge groups for regular and fractional branes respectively. 
One can easily see that in the case of fractional branes 
 the orbifold projects out the numerator of $w_{D3}(\chi,b)$ in (\ref{wd3}) and therefore
 the condition $w_{D3}\approx 1$ cannot be realized in the limit $\epsilon_\ell\to 0$.  
This implies that a probe theory with  $w_{D3}\approx 1$ can be engineered only for regular branes and therefore the reasonings following (\ref{wd32})
only apply in that case.  
   
We consider then a $\Z_2$ orbifold breaking $SO(2N)\times Sp(M) \times SO(2k)$
down to $U(N)\times U(M)\times U(k)$. Notice that unlike in the previous case, now the
theories living on the D7 and on the D3 are both invariant under eight supercharges. 
The instanton moduli space for this model has been worked out in \cite{Billo':2010bd}. 
We refer the reader to this reference for details.

\subsection{The D(-1)D7 system}

We start again by describing the system in absence of D3-branes, whose contributions will be added later. 
The D7 brane theory is now half maximally supersymmetric  and after reduction
 to four dimensions leads to a ${\cal N}=2$  with 
gauge group $U(N)$ and two antisymmetric hypermultiplets.

Including $k$ D(-1) instantons, the symmetry group of the theory becomes  
$U(N)\times U(k)$ with the $(B_{1,2},D_{1,2})$ moduli transforming in the adjoint  
of the $U(k)$. The $B_{3,4}$ transform in the fundamental and anti fundamental representations of the gauge group,
$\Ysymm+\overline{\Ysymm}$, and  $D_{3,4}$ in the antisymmetric and its complex conjugate
$\Yasymm+\overline{\Yasymm}$. 
Finally the D(-1)D7 open strings and their associated ADHM constraint    
transform in the $(\Yfund,\overline{\Yfund})$ and $(\Yfund,\Yfund)$
bifundamental representations respectively of $U(N)\times U(k)$.
  
    The $k$-instanton partition function $Z_{D7,k}$ can be written again as
  (\ref{zd7}), now  with  integrand
     \begin{eqnarray}
 w_{D7} (\chi,a)  &=& {c_k \over k! } \left( \frac{ \epsilon_1+ \epsilon_2  }{\epsilon_1 \epsilon_2 }
\right)^k ~
\prod_{\ell=1}^2 \prod_{i < j}^{k} { (\chi_{ij,-}^2-s_\ell^2)(  \chi_{ij,+}^2-s_{\ell+2}^2 )
 \over
  ( \chi_{ij,-}^2-\epsilon_\ell^2  )   ( \chi_{ij,+}^2-\epsilon_{\ell+2}^2  )    }\nn\\
&& \times  
  \prod_{\ell=1}^2 \prod_{i=1}^k    \frac{1}{\Big(4\chi_i^2
-\epsilon^2_{\ell+2} )    }
\prod_{i=1}^k\prod_{u=1}^{N} \Big(\chi_i-a_u\Big) ~   \label{wd72}
\end{eqnarray}
    $c_k=(-)^k$ and $\chi_{ij,\pm}=\chi_i \pm \chi_j$.  
      The eight dimensional prepotential and chiral correlators are defined as before 
  via  (\ref{ff4}) and (\ref{chiral}). 
   
\subsection{The coupling on the D3-brane probes}

Now let us consider the theory on a D3-brane probe.  The theory on the probe is
${\cal N}=2$ supersymmetric with gauge group $U(M)$. 
The matter content  
 is given by two hypermultiplets transforming in the antisymmetric representation and $N$ 
chiral multiplets transforming in the fundamental representation of $U(M)$.

The beta function coefficient becomes
\footnote{The index of $U(N)$ representations are 
 $T({\rm adj})= N$~,~~~
$T(\Yasymm) =\frac{N-2}{2}$, 
$T(\Yfund)=\frac{1}{2}$.} 
  \bea
b_{U(M)} &=&  4 - N 
\eea
 As before the case $N=4$ is special in the sense that the four dimensional gauge theory is conformal. 
 
Now let us consider the contributions of D(-1)-instantons.  The instanton partition function
$Z_{D3,k}$  can be written again as (\ref{zd3}) but now with $w_{D7}$ given by (\ref{wd72})
and    \cite{Billo':2010bd}
  \be
  w_{D3} (\chi,b) = \prod_{i=1}^k \,
\prod_{m=1}^{M}
\frac{\Big(( \chi_i
+b_m)^2-\frac{(\epsilon_3-\epsilon_4)^2}{4}\Big)}{\Big((\chi_i
-b_m)^2-\frac{(\epsilon_1+\epsilon_2)^2}{4}\Big)}   \label{wd3u}
 \ee 
   The four dimensional prepotential and gauge couplings are defined as before by 
   (\ref{fd3}) and (\ref{taumn}). 

 Now we would like to relate the gauge coupling on the probe to the chiral ring of the D7 brane theory.  Trying to follow the same strategy as 
before, we immediately run  into problems
 because the function $w_{D3}$ does not fall to one at $\epsilon_\ell \to 0$. This can be cured
 by choosing the regular D3-branes symmetrically distributed around the origin 
  \be
b_m =\left\{     
\begin{array}{lll}
 (b_n,-b_n) &  {\rm for} & M~{\rm even}  \\
 (b_n,-b_n,0) &  {\rm for} & M~{\rm odd}  \\\end{array}
\right.   \qquad    n=1,\ldots, [\ft{M}{2}]   \label{chis2}
\ee  
 For this choice one finds  
 \be
w_{D3} \,(\chi,b)=1+2 \epsilon_3 \epsilon_4 \sum_{J=0}^\infty \sum_{n=1}^{[M/2]}   
{{\rm tr} \chi^{J} \over b_n^{J+2}   }  (J+1)+\ldots    \label{wd32p}
\ee
  Following the same steps as in the $SO(2N)$ case one is left with
 \bea
 2\pi i  \, \tau^{\rm inst}_n = 2  \left \langle   \ln {\rm det} \left(1- {\Phi\over b_n} \right) 
  \right  \rangle_{ {\rm inst}}    \label{taumn3}
\eea
As before, the elementary $SU(2)$ probe, i.e. $M=2$, captures the full information about the axion-dilaton background.  
It is important to notice that for the choice (\ref{chis2}) only correlators involving  even powers of $\Phi$  are non-vanishing. 
Moreover, by an explicit evaluation of the eight-dimensional chiral correlators, one finds that the results are just half of those 
for the $SO(2N)$ case leading to the same answer for the gauge coupling $\tau_m$ of the elementary $SU(2)$ probe.
This is not surprising since the two probe theories share their $SU(2)$ content since 
  the antisymmetric  representation of $SU(2)$ is the trivial in this case.

\section{The axion dilaton profile}

 The operator $ {\cal O}_{\tau} (b,q,\Phi)$ entering in the eight dimensional chiral correlator
 found before can be alternatively derived from string disk diagrams describing the 
 rate of emission of a dilaton-axion field from D7, O7 and D(-1) sources. 
 This has been done in \cite{Billo':2011uc} for the SO(8) case but it
holds in the more general settings under consideration here. 
In this section we review these results and comment on the generalization to the orbifold
case.

The axion-dilaton profile emitted by a D-brane source is computed by a disk involving 
a closed string vertex for this field and any number of open string insertions. 
For concreteness let us focus on the dilaton field  with string vertex
\be
V_{\phi} =\delta\phi(p) \, c \tilde c \, e^{-\varphi-\tilde \varphi}\, \Psi^M \,   \tilde\Psi_M\,  e^{ip \cdot X} 
\ee
where $\delta\phi(p)$ is the field polarization, $\Psi, X$ fermionic and bosonic fields and $c, \varphi$ ghosts and superghosts.
As usual the tilde indicates fields of opposite handedness.
 We consider a dilaton profile along the two dimensional plane transverse to the D7 branes
 and therefore we take the momentum $p$  of the closed string vertex along this plane.   Denoting by $a$ and $\chi$  
the position matrices of the D7 and D(-1) instantons respectively along this plane, the effective coupling of the dilaton to 
the D(-1),D7 sources are given by the disk amplitudes
\bea
  \left\langle\,   V_{\phi}  \, e^{-{i a\over 2\pi}  \int d\tau \partial_\tau X }  \right\rangle_{{\rm disk}, D7} & \sim &  \delta\phi\,  {\rm tr} 
\, e^{ip \cdot a} \nn\\
\left\langle\,   V_{\phi}  \, e^{-{i \chi\over 2\pi}   \int d\tau \partial_\tau X } \right \rangle_{{\rm disk},D(-1)} & \sim &  \delta\phi\, {\rm tr}
\, e^{ip \cdot \chi}   \label{disks}
\eea
In presence of an O7 plane one finds an extra contribution 
$ \left\langle\,   V_{\phi}  \,   \right\rangle_{O7}  \sim   -8 \delta\phi $ and the amplitudes are projected
onto their even parts under reflections i.e. the matrices 
$a$ and $\chi$ are projected into the adjoint of $SO(2N)$ and $SO(K)$ groups respectively. 
Similar couplings are found for the axion field $C_0$, that  
 results in the replacement $\delta\phi \to \delta\tau$  in (\ref{disks}) with $\tau=e^\phi+i C_0$ 
 the axion-dilaton field.
 After Fourier transforming one can write the D7 contribution to the axion-dilaton effective action as
  \bea
  \delta S_{\rm eff,D7} &=&
 -2\pi i \,  \int d^{8} x     \, {\rm tr} \, e^{a\cdot \partial_z}  
  \delta \tau(z) +{\rm h.c.}   \label{seff7}
  \eea
  On the other hand the coupling  to the D-instanton modifies the instanton moduli space by
   \bea
         \delta S_{\rm mod} &=&  - 2\pi i \, {\rm tr} e^{\chi \cdot\partial_z} \delta \tau(z)  +{\rm h.c.}  \label{seffs1}
    \eea
  The contribution of this term to the effective action can be computed by the standard localization
  formulae
  \bea
   S_{\rm eff,D(-1)} &=&\int d^8 x d^8   \theta\,   {\cal F}_{D7}(M,T)   \label{sfmt}
  \eea
 with  ${\cal F}_{D7}(M,T)$  the eight dimensional prepotential
  \bea
   {\cal F}_{D7}(a,\delta\tau)
  &=& -\lim_{\epsilon_\ell\to 0}  \epsilon_1\epsilon_2\epsilon_3\epsilon_4 
       \ln Z_{D7}(a,\delta\tau)
  \eea
 given in terms of the multi instanton partition function
\be
 Z_{D7}(a,\delta\tau)=1+ \sum_{K=1}^\infty q^K \int d{\cal M}_K e^{-S_{\rm mod}({\cal M}_K,a)-   \delta S_{\rm mod} (\delta\tau) } 
\ee
 Finally
 \bea
 M &=&\Phi+\ldots \nn\\
 T &=& \tau+\ldots +\theta^8 \partial^4 \bar \tau   \label{superfields}
 \eea
 are the superfields containing the chiral field $\Phi$ on the D7-brane and the axion-dilaton field respectively.
To linear order in $\delta T$ one finds
\bea
\delta{\cal F}_{D7}(M,\delta T) &=&   -2\pi i \,  \sum_{J=0}^\infty  \frac{\partial^{J}  \delta T }{J!} 
\lim_{\epsilon_\ell\to 0} { \epsilon_1\epsilon_2\epsilon_3\epsilon_4
\over Z_{D7}} \nn\\
&&\times \sum_K q^K  \int  d {\cal M}_K\, e^{-S_{\rm mod}( {\cal M}_K,M  )  }  \,     \, {\rm tr} \,  \chi^{J} \nn\\
 &=&     -2\pi i \,   \sum_{J=0}^\infty  \frac{ \partial^{J}  \delta T }{(J+4)!}  
   \left \langle  \, {\rm tr} \,   \Phi^{J+4}   \right\rangle_{ {\rm inst}}
   \label{fchi}
\eea
 where in the second line we used (\ref{chiral}) to relate the integrals in  right hand side 
 to 
the chiral correlators in the D7 gauge theory.  Plugging this into (\ref{sfmt}) and using 
the highest component of $T$ in (\ref{superfields}) to soak the $\theta$-integrals one 
finally finds
 \bea
 \delta S_{\rm eff,D(-1)}  
   &=&  - 2\pi i \,  \int d^{8} x     \left \langle  \, {\rm tr} \,  e^{\Phi\cdot \partial_z}   \right\rangle_{ {\rm inst}}
  \delta \tau(z) +{\rm h.c.}  \label{susyr} \eea
   Notice that (\ref{susyr}) is nothing but the non-perturbative analog of 
 the coupling  $S_{\rm eff,D7}$ found in (\ref{seff7}) and therefore one can write the two together as
\bea
  \delta S_{\rm eff}  & =& \delta S_{\rm eff,D7}  + \delta S_{\rm eff,D(-1)} \nn\\
  &=&
   -2\pi i \,   \int d^{8} x     \,\langle  {\rm tr} \, e^{\Phi \cdot \partial_z}   \rangle_{D7} 
  \delta \tau(z) +{\rm h.c.}  
 \eea
with
    \bea
  \langle {\rm tr} \Phi^{J} \rangle_{ D7}  = 
  {\rm tr} \,a^{J}     + \langle {\rm tr} \Phi^{J} \rangle_{ {\rm inst}}  
   \eea
   denoting the full chiral correlator in the eight dimensional theory with the first term
   accounting for its classical  part. Taking into account the bulk kinetic term 
\be
S_{\rm bulk} =\int d^{10} x \, \bar{\tau }\, \partial \bar{\partial}\, \tau 
\ee
 and varying with respect to $\bar \tau$ one finds the equation of motion
  \be
  \partial_z \bar \partial_z \tau = 2\pi i\, \left[ \sum_{J=0}^\infty  \ft{1}{J!} \partial_z^{J}   \delta^{(2)}(z) \, 
  \langle {\rm tr}\, \Phi^J \rangle_{D7}   -8  \delta^{(2)}(z) \right] 
  \ee
 that is solved by
 \be
 2\pi i \, \tau(z) =2\pi i  \,\tau_0 + \left \langle   \ln {\rm det} \left(z-\Phi  \right) 
  \right  \rangle_{ D7}   -8 \ln z  \label{tauz}
  \ee
  The harmonic part in (\ref{tauz}) has been fixed by matching the one-loop result on the two sides of the equation.   
The result (\ref{tauz})  agrees with the gauge coupling found from the direct multi-instanton computation (\ref{taumn2}) on the D3-brane probe with $z\to b_m$. 
  The generalization to the orbifold case is straightforward. We notice that the orbifold acts as a projection on the open string degrees of freedom   
and therefore all formulae in this section still apply with the only difference that now the  moduli space integrals defining the 
chiral correlators run over the  $\Z_2$-invariant moduli subspace.

\vskip 1cm \noindent {\large {\bf Acknowledgments}} \vskip 0.2cm

We thank M.Bill\`o, M.L.Frau, A.Lerda and L.Martucci for useful discussions.
This work was partially supported by the ERC Advanced Grant n.226455 {\it ``Superfields''}, by the
Italian MIUR-PRIN contract 20075ATT78, by the NATO grant
PST.CLG.978785.

\noindent \vskip 1cm

  \begin{appendix}

 \section{Eight dimensional correlators}

In this appendix we collect the results  for the eight-dimensional correlators. 
We start by considering the case of D7-branes in flat ten-dimensional space realizing 
an $SO(2N)$ gauge theory in eight-dimensions. 
The correlators follow from explicit
evaluation of the integrals in (\ref{chiral}).  
 Up to order $q^3$  one finds the non-trivial correlators \cite{Fucito:2009rs}
  \footnote{With
respect to  \cite{Fucito:2009rs}  we perform the rescaling $q\to -\frac{q}{16}$. } 
      \bea
      \langle {\rm tr} \Phi^{4} \rangle_{\rm inst}  &=&  -12 \, q \, \sqrt{{A}_{N}}-\frac{3 \, {q}^{2} \, {A}_{N-2}}{8}-\frac{3}{16} \, {q}^{3} \, {A}_{N-4} \, \sqrt{{A}_{N}} \nn\\ 
      \langle {\rm tr} \Phi^{6} \rangle_{\rm inst}& =& \frac{45 \, {q}^{2} \, {A}_{N-1}}{8}+\frac{15}{8} \, {q}^{3} \, {A}_{N-3} \, \sqrt{{A}_{N}}\nn\\
         \langle {\rm tr} \Phi^{8} \rangle_{\rm inst}  &=&-\frac{105}{4}  \, {q}^{2} \, {A}_{N}-\frac{35}{4} \, {q}^{3} \, {A}_{N-2} \, \sqrt{{A}_{N}}\nn\\
         \langle {\rm tr} \Phi^{10} \rangle_{\rm inst}  &=&\frac{\hbox{105}}{4} \, {q}^{3} \, {A}_{N-1} \, \sqrt{{A}_{N}}\nn\\
\langle {\rm tr} \Phi^{12} \rangle_{\rm inst}  &=&-\frac{495}{8}  \, {q}^{3} \, {A}_{N}^{3/2} \label{8corr}
     \eea
    with  $A_m$, $m\,=\,1,\ldots,N$ given  in (\ref{ans}). 
    
    In the case of $U(N)$ plugging (\ref{wd72}) into (\ref{chiral}) one finds that the results for the
    correlators are half of those in (\ref{8corr}).

 \end{appendix}

\newpage

\providecommand{\href}[2]{#2}\begingroup\raggedright\endgroup

\end{document}